# Female Teenagers and Coding: Create Gender Sensitive and Creative Learning Environments

**Bernadette Spieler,** *bernadette.spieler@ist.tugraz.at*
Institute of Software Technology, Graz University of Technology, Austria

**Wolfgang Slany,** *wolfgang.slany@tugraz.at*
Institute of Software Technology, Graz University of Technology, Austria

## Abstract

The number of women in technical fields is far below the average number of males, especially in developed countries. Gender differences in STEM are already present in secondary schools in students aged between 12 to 15 years. It is during this intermediate female adolescence that girls begin to make critical career choices, which therefore makes this a key age to reinforce them and reduce the gender disparities in ICT. Acquiring computational thinking (CT) skills, particularly coding, is important for building a positive economic, developmental, and innovative future. To address the gender bias in schools, one of the goals of the European H2020 project No One Left Behind (NOLB) included integrating Pocket Code, a free open source app developed by the non-profit project Catrobat, into different school lessons. Through game design, Pocket Code allows teenage girls to incorporate diversity and inclusiveness, as well as the ability to reflect their cultural identity, their likes, and their ways of interacting and thinking. To evaluate the impact of the use of the app in these courses, we captured the results on engaging girls in design and coding activities. For this paper, the authors present the data of surveys via a qualitative content analysis during the second cycle of the project. The results let the researchers conclude that the organization and the setting of the coding courses (for example, guidance and supporting material, freedom of choice) had much more influence on female students' engagement than the coding aspects or the app itself. In contrast, male students more frequently mentioned missing features in the app, and stated that they liked the coding. With a focus on female teenagers, the results allow us to conclude that a suitable classroom setting is significantly more important for them than the coding tool itself.

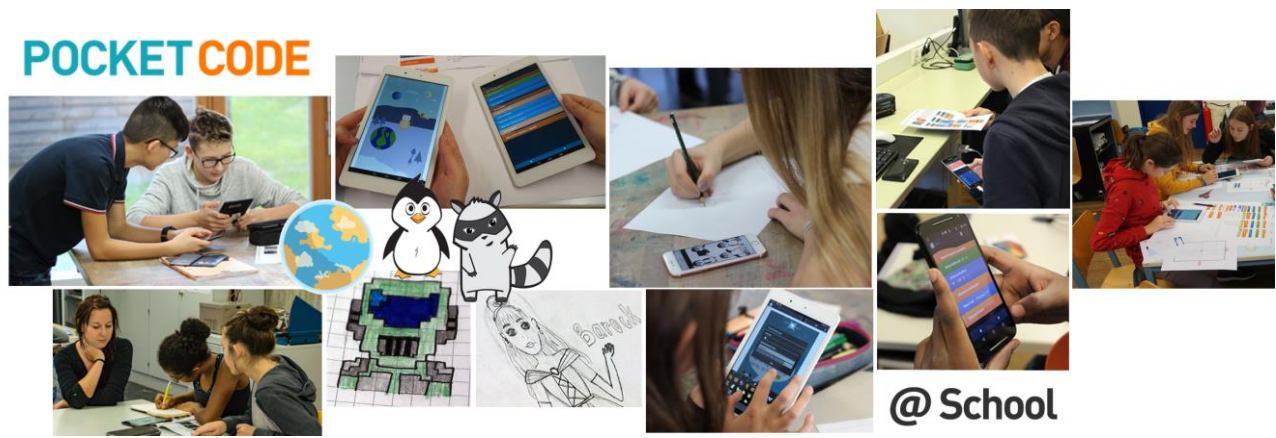

*Figure 1: Providing inclusive coding environments for female teenagers*

## Keywords (style: Keywords)

Pocket Code, Game Design, Gender Inclusion, Coding, Mobile Learning, STEM, Social Inclusion, Constructionism, Girls, Learning Environment

## Introduction

Secondary school is the place where students make the critical choices which decide their future careers, develop a more realistic picture of their future jobs, and assess their career-relevant abilities. Researchers observe that girls' interest in IT drops significantly from the age of 12 to 15 (Tsan, Boyer, and Lynch, 2016). To address this gender bias at an early stage, a goal of the European project No One Left Behind (NOLB) was to integrate Pocket Code, an app developed at Graz University of Technology, into different school subjects, thus making coding more accessible and attractive to female students. Although promoting gender equality is a longstanding policy which all European countries place on their agendas (EC, 2014), a gender-based inequity in ICT still poses barriers for women. Thus, the acquisition of digital skills is more important than ever and represents a key professional qualification (Balanskat and Engelhardt, 2015, Kahn, 2017, Tedre and Denning, 2016). For the NOLB project, the authors assumed that according to related literature it is possible to spark girls' interests by getting them engaged in computational thinking through collaborative, creative, and engaging coding activities (Khan and Luxton-Reilly, 2016). Thus, the team studied whether the design of mobile games has an impact on inclusion and satisfaction of female students. Thus, the following research question has been defined: How can we organize coding activities to reinforce female teenagers in computer science?

The paper is organized as follows: First, we describe the challenges of teaching CS in Europe and discuss several learning theories, with a focus on constructionism. Second, we present solutions on how to provide suitable learning environments for girls, followed by a brief description of the NOLB project and our educational apps for coding. Third, we present the results of the quantitative and qualitative surveys which show insights for framing suitable classroom settings for coding activities for girls. Finally, the last sections conclude the paper and describe our future work.

## Learn Coding: A Worldwide Challenge

The European Commission states, "*All of Europe's citizens need to be educated in both digital literacy and informatics*" (Informatics Europe/ACM Europe, 2013, EC, 2016). Thus, IT education is seen as an interdisciplinary field that bridges the gap between the use of digital media and information-processing technology as well as basic concepts and fundamental ideas of computer science. The EC recommend that students should benefit from computer literacy at an early stage and that it is important to find a standardized definition of the informatics curricula through all countries. However, in Austria and in many other European countries, computer science topics are underrepresented in school curricula, hence, teaching time for these topics is limited (Informatics Europe/ACM Europe, 2016). From primary through secondary school, only a few opportunities exist for students to explore coding and CS topics.

Learning theories from the past serve as an organized set of principles, clarifying how people acquire, retain, and recall knowledge (Schunk, 2016). These theories helped researchers to gain a better understanding of how learning occurs and help us to select appropriate techniques, tools, and strategies to support learning and teaching how to code. Three basic types of learning theory exist: Behaviorism, Cognitivism, and Constructivism, and some subtypes or variations, e.g., Instructionism, and Constructionism.

Figure 2 provides an overview of important findings of each theory, how each theory handles students' motivations, and shows details on how teaching and learning should be done.

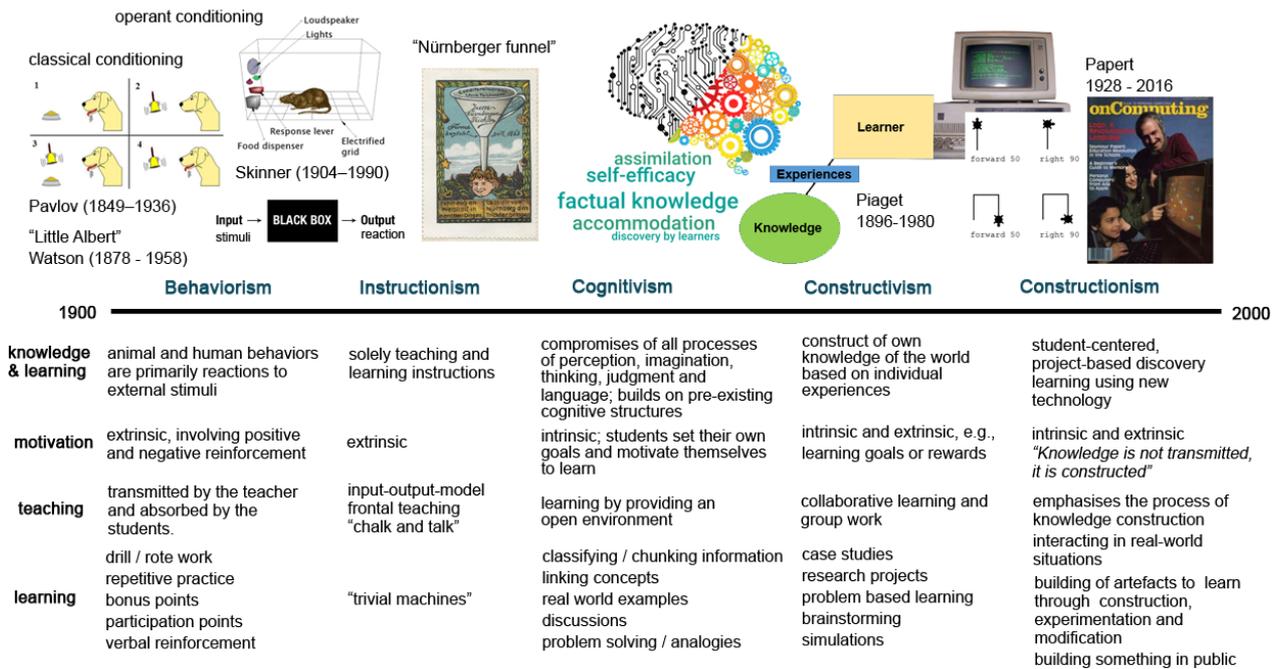

*Figure 2: Learning theories of the 20th century: Behaviorism/Instructionism (Pavlov, 1927, Skinner, 1976, von Foerster et al., 2009), Cognitivism (Piaget, 1968, Perry, 1999), Constructivism (Piaget, 1968, Vygotsky, 1978), Constructionism (Papert, 1980, Paper, 1991).*

## Constructionism

The Constructionist approach (Papert, 1980) is interested in building knowledge through active engagement and personal experience. Papert noted that individual learning occurred more effectively when students understood the world around them and were creating something that was meaningful to them. This experiential and discovery learning by challenges should inspire creativity, and project work allows for independent thinking and new ways of constructing information. The iterative process of self-directed learning underlines that humans learn most effectively when they are actively involved in the learning process and build their own structures of knowledge. In this theory, communication between students about the work, and the process of learning with peers, teachers, and collaborators, is seen an indispensable part of a students' learning (Papert, 1993, Papert, 1991).

> *"The construction of knowledge through experience and the creation of personally relevant products. The theory proposes that whatever the product, e.g. a birdhouse, computer program, or robot, the design and implementation of products are meaningful to those creating and that learning becomes active and self-directed through the construction of artefacts."* [Papert, 1980, p.2]

Thus, Papert described the huge potential of bringing new technology into the classroom (Papert, 1993). For this reason, he co-invented the LOGO programming language in the late 1960s at the MIT. LOGO was designed to have a "*low threshold and no ceiling*" and was indeed used to help novice programmers, and to support complex explorations and the creation of sophisticated projects (Tinker and Papert, 1989). LOGO set the basis for later visual programming tools, such as Etoys (Kay *et al.*, 1997) or Scratch (Resnick *et al.*, 2009). Such block based visually oriented tools made programming accessible for a large number of people and taught new skills such as engineering, design, and coding (Blikstein and Krannich, 2013). They allow students to recognize blocks instead of recalling syntax. They are broadly integrated in schools, or even at universities all over the world (Meerbaum-Salant et al., 2010).

To conclude, psychologists and pedagogues from today following the constructionist approach state three main goals. First, they wish to rethink traditional education without step-by-step guidance and to create new social and open environments. Second, they strive to allow students to engage in meaningful and relevant problem-solving activities, and third, they want to integrate new tools, media, and technologies in school lessons (Neo and Neo, 2009).

Jeannette Wing, 2006 shaped the term "Computational Thinking" (CT):

> *"Computational thinking involves solving problems, designing systems, and understanding human behavior, by drawing on the concepts fundamental to computer science" (Wing, 2006)*

Wing's idea that children who are introduced to CS learn more than just programming opened a new way of thinking, e.g., it showed the benefits of learning to think like a technician (Wing, 2008). Wing's findings have been incorporated into the CS curriculum of many countries (Kahn, 2017) and into K-12 movements (Mannila et al., 2014). However, CT is just a small subset of Seymour Papert's ideas in the 80s (see previous section). Papert was the first who used the phrase computational thinking and defined it in a much broader way. For instance, Wing focuses mainly on computer programs, whereas Papert stated that there are more kinds of constructionist projects, and computational ideas could serve learning in a broad variety of subjects, this "*can change the way* [children] *learn everything else*" (Papert, 1980, p. 8). In comparison, CT concepts lack creativity and student-directed projects (Kahn, 2017). To summarize, CT concentrates on the importance of coding and computer science activities, thus delivering concepts that are more applicable and successful on a number of levels (Tedre and Denning, 2016). However, critics argue that coding should not be seen as a unitary skill but instead as a meta-skill for a complex network of other skills.

## Creative Environments to Reinforce Female Teenagers in Coding

Teaching good game design and development skills is especially important for girls because they are not that likely to play games (Krieger, Allen and Rawn, 2015). In addition, the literature argues that even if the number of female students who plays video games increased in recent years, male students have a greater interest in playing games (Jenson, Castell and Fisher, 2007). However, a recent study which examined American female players' experiences showed that 65% of women play mobile games, compared to 2011 when only 31% of women played mobile games (Google and NewZoo, 2017). Furthermore, 64% of women prefer smartphones to other platforms (38% of the men do so). Thus, to use smartphones to design mobile games seems to be a very promising approach to attract young females. Furthermore, framing a supportive classroom setting for coding activities is a critical factor, and literature suggests that interventions should specifically target the classroom climate to strengthen teenage girls' confidence to motivate female students extrinsically as well (Beyer *et al.*, 2003). Such first positive experiences in coding may direct their future career choices towards STEM fields. If they become game designers and creators of their own learning content through a constructionist, creative, and engaging learning environment, this can significantly contribute to closing the divide and participation gap in digital culture (Veilleux *et al.*, 2013, Allison, Cheryan, and Meltzoff, 2016). Thus, the literature states that for young women, creativity and interest in STEM professions are often related, but there is a lack of practical relevance in a lot of these subjects. A European-wide study in which 11,500 young women between 11 and 30 were interviewed (Microsoft, 2017) showed that girls between the ages of 12 and 16 are the most creative. Approximately every third women that has been asked (33%) criticized how scientific topics were explained in schools and that those subjects are taught from a more "male perspective". Thus, it is important to make CS more attractive for girls in order to sustain a balance in this very strongly male-dominated IT labor market. It is therefore important to eliminate the prejudice that STEM professions are not creative. Through game design activities, creative environments, and customization, a broader spectrum of girls can be reached (Subsol, 2005). If facilitators promote especially IT careers that are more driven by creative thinking and design, more female students will expressed their interest in IT careers (Wong and Kemp, 2017).

# The European No One Left Behind (NOLB) Project

The focus in the subsequent sections lies on the European No One Left Behind[1] (NOLB) project and the Catrobat[2] learning apps, Pocket Code and Create@School. The NOLB project has been funded by the Horizon 2020 framework and involved partners from Germany, Spain, the UK, and Austria. The vision of the NOLB project was to unlock inclusive game creation and to construct experiences in formal and informal learning situations from primary to secondary level, particularly for students at risk of social exclusion. This project started in January 2015 and reached its conclusion in June 2017 by validating its outcome in different phases: preparation phase, feasibility study, first, and second cycle. To limit the scope of topics to those relevant to this paper, the remainder of this section focuses on the results of the *second cycle* and on the *Austrian pilot*. In the past the authors' work concentrated on the feasibility study (Petri *et al.*, 2016), evaluation of performed Pocket Code Game Jams (Spieler *et al.*, 2016), and analyzing female teenagers' performance in regard to the learning goal achievement of submitted programs (Spieler, 2018).

## Educational Coding Apps

Our app Pocket Code (Slany, 2014), is a visual programming language environment that allows the creation of games, stories, animations, and many types of other apps directly on smartphones or tablets, thereby teaching fundamental programming skills. Programs in Pocket Code follow a similar syntax to the one used in Scratch. During NOLB an enhanced version of the app has been developed and adapted for use in schools. This new version is called Create@School and integrated the results of the observations during the pilot studies as well as feedback from teachers and students. Furthermore, a web based Project Management Dashboard (PMD) for teachers was developed. The app was released in October 2016 for first test runs during the second cycle of the project. Important components of the new Create@School flavor of Pocket Code were the gathering of analytics data, the integration of accessibility preferences for children with special needs, and pre-coded game design templates. In previous work the authors focused on the evaluation of the Create@School features (Spieler *et al.*, 2017 [1]) and the teachers' perspective during NOLB (Spieler *et al.*, 2017 [2]). This paper focuses on the evaluation of the NOLB coding environments, thus the apps will not be explained in more detail.

In Austria, the NOLB project piloted in three different schools, in total, 478 students participated in Austria (281 female students and 197 male students). Altogether, 22 coding courses were conducted, and the coding apps were integrated in the curricula of English, computer science, physics, fine arts, and music (Spieler *et al.*, 2017 [2]).

## Gender Inclusiveness in NOLB

The idea of NOLB included that game creation challenges in classes should enhance female students' abilities across all academic subjects, including logical reasoning, creativity, and the development of social and computational thinking skills (see previous section). Moreover, students had the opportunity to socialize with their peers during the game making process by working in teams. To address the gender bias in coding classes, the goal of the Austrian study included how to make the coding environment more suitable for female teenagers. From the literature, the assumption for NOLB was to attract girls by:

- Increasing their personal attachment to programming by improving our services with appropriate example games/templates, new assets, and themed tutorials/templates.
- Increasing their involvement by encouraging them to become active members of the Pocket Code community by providing them a safe and interesting environment to join, with featured games from female users for female users.
- Asking them to design their own games rather than just to code programs.

---

[1] http://no1leftbehind.eu/
[2] https://www.catrobat.org/

The intent was to discover how to organize the coding units in ways that specifically empower girls by engaging them with playful and creative activities.

## Evaluation of NOLB Activities in Austria

In this section, the NOLB results from the Austrian case study are evaluated to investigate female students' experiences, behaviors, and outcomes when using our tools and services in different courses. With the help of a quantitative and qualitative survey, students' opinions about the user experience (UX) of the Create@School app and the courses in which the app was used has been collected to gain a deeper understanding of the experience evaluation. Overall, a total of 131 students filled out feedback forms in Austria during the second cycle (63 male and 68 female students). Based on the research design, a t-test was performed which compares whether two groups (female/male students) have different average values. The evaluation is part of the NOLB Delivery 5.4 (Spieler and Mashkina, 2017). Results are collected by answering like/dislike questions: *How was your experience with Create@School? (very good – very bad).* Figure 3 illustrates students' opinions per answer in percentage. The percentage of female students (in orange) who rated the experience as "good" was 29% (38), and the percentage of male students (in cyan) account for 21% (27). The answer option "bad" was chosen by 13% of the girls (17) and 14% of the boys (19). In general, girls rated the app experience more positively (mean female = 2.63) as their male colleagues did (mean male = 2.37, but not significantly: $p$ = .130, $α$ = .05).

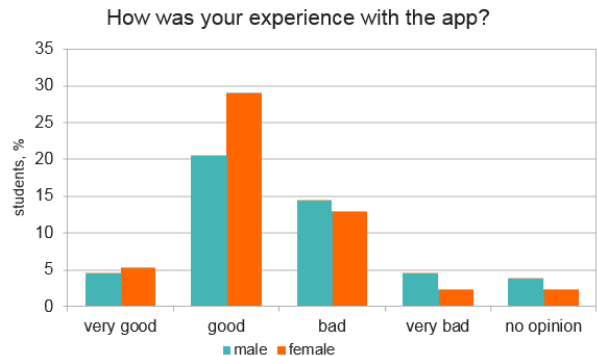

*Figure 3: Distribution of answers about the experience with the Create@School app.*

To clarify the motivation for these answers, it is necessary to take a closer look at the analysis of the open-ended questions: *What did you like the most?* The answers of the participants describe their positive impressions about the Create@School experience. Answers could be classified into five different categories: "working process", "the app", "the results (their game)", "organization", or "others". The distribution of the answers among the categories can be seen in Figure 4. This time a Mann-Whitney U test was performed to analyse whether the central trends of the two independent samples are different.

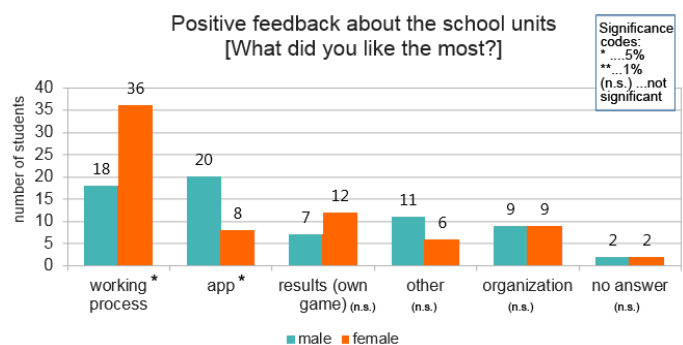

*Figure 4: Categorization of the positive impressions during NOLB.*
*Note that one answer could contribute to two categories, for instance the feedback "the finished products + the facilitators" contributed to both, results and organization categories.*

The category "working process" contains feedback about preferred actions (e.g., I like to program, to play the game, to design), or properties of the school units (e.g., having a freedom of choice, that you could be very creative). The largest statistically significant difference between the responses of male and female participants was observed for this category, namely 26% of the girls (36) and 13% of the boys (18) provided positive working process related feedback ($z = 2.402$; $p = .016$, $α = .05$). The category "app" contains the answers mentioning the experiences directly connected with the app itself, e.g., "the simplicity of the app", or "the different effects and backgrounds". Male students seemed to be significantly more positively impressed by the app than girls were ($z = -1.969$; $p = .049$, $α = .05$). A number of 8 girls and 20 boys evaluated the app qualities as positive. Some students were especially satisfied with the results of their work or the concept of game creation, and the category "results (own game)" contains this type of feedback. For instance, "the results and how everything turned out", or "I liked the idea of creating a game" demonstrates this feedback. Of the girls, 9% (12) and 11% of the boys (7) were satisfied with the results or their personal game. The answers from the category "organization" highlights feelings about how the unit was structured. This includes if students were enjoying teamwork or solving the problems on their own, the presence of external people (facilitators of the workshop), usage of tablets during the school units, etc. It was noticeable that students in the younger age group of 12 - 14 years old were very excited about using the tablets during the school units. Typical responses for this category were, e.g., "to work in a team with your friends", "when you (the facilitators) explained to us how the app really worked", "that we had our own tablets". The same number of the male and female participants' feedback falls into the category of "organization" with nine responses from boys and from girls. The replies that could not be clearly classified into any of the described categories above were summarized into the "other" category. For example, answers with no clear message, or containing feedback connected to a particular game or classroom setting, as well as such responses as "nothing" or "everything".

Next, we will take a closer look on the categories "working process" and "app". The category "working process" was divided into six subcategories: "designing", "creativity", "playing", "freedom of choice", "programming", and "other". Figure 5 shows the distribution of the responses.

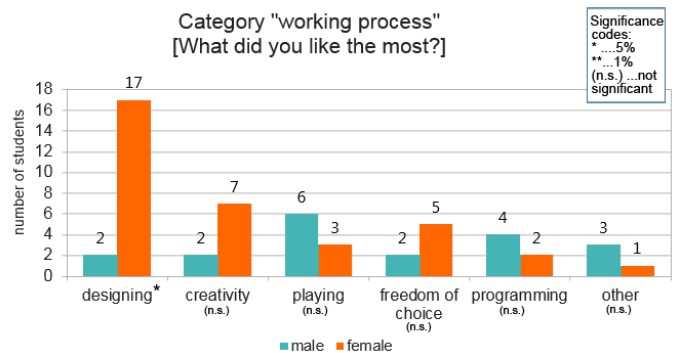

*Figure 5. Detailed overview of positive impressions category "working process"*

The subcategory with the name "designing" summarizes the feedback related to drawing, taking pictures, personalization of the characters, etc. For instance, "drawing our pictures" or "creating the characters" fall into this subcategory. There were 32% of the girls (17) and only two boys who identified design related activities as the essence of their positive experience (significant: $z = 2.151$; $p = .032$, $α = .05$). The subcategory "creativity" represents the feedback praising the creative side of the school units, for instance, "you could be very creative", "you could do almost any game you wanted". A total of seven female students and two male students evaluated creativity as the positive aspect of the Create@School experience. A total of six boys and three girls stated that they considered "playing" as a positive experience. Answers like "you were allowed to play" were typical for this subcategory. No constraints in choice or actions were valued by two boys and five girls. These responses were summarized within the subcategory "freedom of choice". A representative statement for this subcategory was "That you could do whatever you want". There were four boys and two girls who enjoyed the programming process itself. Typical responses for the subcategory "programming" were "programming the rocket" or "programming by ourselves".

The category "app" was divided into four subcategories: "LEGO®-style bricks", "features", and "design". The distribution of the feedback can be seen in Figure 6.

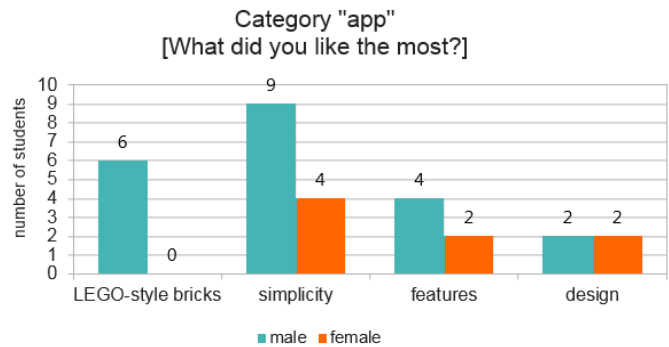

*Figure 6: Detailed overview of positive impressions of the category "app".*

None of the girls but six boys stated that they were pleased about the LEGO-style bricks in Create@School. Although the number of answers is too small to be significant, the insights are interesting. The response "brick system" was representative for this subcategory. A total of nine boys and four girls liked the "simplicity" of the app. Feedback included "It was relatively self-explaining!", or "components are easy to understand". The subcategory "features" consists of answers like "the different effects and backgrounds" or "variables were available". Four boys and two girls contributed to this subcategory. Equally, two of each gender liked the design of the app.

*What did you like the least? Any suggestions for app improvement?* The answers to the questions were the basis for the negative feedback evaluation. Note that some students responded with everything was fine (6) and some did not give any answer to this question (10). Figure 7 categorizes the negative impressions about the app and the school units.

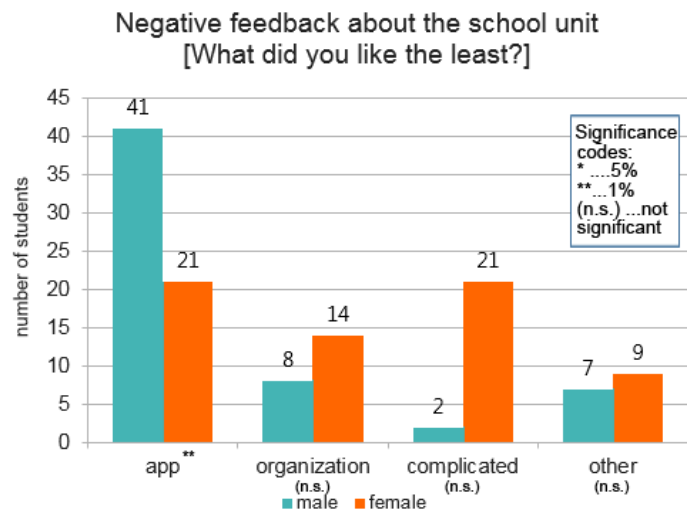

*Figure 7: Categorization of the negative impressions about the app and the school unit. Note that one answer can contribute to two categories, for instance the feedback "It was too complicated and it took too long to finish the game" contributed to both the "organization" and "complicated" categories.*

The category "app" consists of the feedback related only to the Create@School app, for example, "it looked kind of tacky, complicated, confusing", "the axis of the screen", or "it was buggy". There were 33% of the boys (41) and 17% of the girls (21) who provided this kind of feedback (significant: $z$ = -3.372; $p$ = .0007, $α$ = .01). Responses such as "it took too long to finish up the game", "introduction of the app was boring", or "The instructors were not able to explain everything to us" about the unit were summarized into the category "organization". There were 8 boys and 14 girls who gave this kind of responses. The feedback of the contents of the type "I found some things complicated", "make it simpler for people that are not technical", or "you needed help a lot" were summarized under the category "complicated". There were 21 girls (17%) and only 2 boys who gave this type of feedback about their experience with Create@School. The replies that could not be clearly classified into any of the categories described above were summarized into "other", for example, the answers with no clear message, or complaints about the devices and other apps. A breakdown of the category "app" into subcategories can be seen in Figure 8.

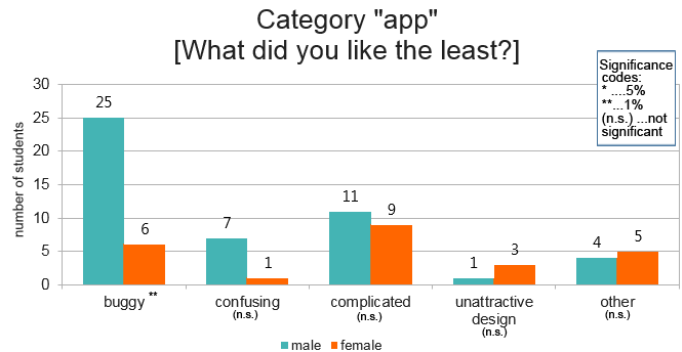

*Figure 8. Detailed overview of negative impressions of the category "app".*

The subcategory "buggy" contains the feedback about the app behavior that students considered as bugs, responses about slow performance, and mentions of app crashes. 32% of the boys (23) and only 8% of the girls (6) gave this type of feedback about the app (significant: $z$ = -2.73; $p$ = .006, $α$ = .01). The representative answers for this subcategory were "Create@School crashes ALL the time", "It sometimes stopped, didn't always work smoothly", or "it took AGES to load". Seven boys and one girl stated that the app lacks structural clarity, and is confusing, e.g. "the app totally lacks the structure", or "some things should be easier to find". These answers were summarized within the subcategory "confusing". Nine girls and eleven boys complained about the complexity of the app (subcategory "complicated"). Comments characterizing the subcategory "unattractive design" were, for example "It looked kind of tacky" or "the design is rather boring". The subcategory "other" contains all other responses that were related to the app, e.g., missing features ("the axes of the screen should be like in Scratch").

## Conclusion and Discussion

The results of the UX experiences with Create@School showed many significant responses and helped the authors to shape future workshops. Overall, the answers about likes and dislikes can be clearly separated by gender. On the one hand, female students significantly preferred the aspects of the Create@School units ("working process") and mostly related to design activities, but did not mention the programming aspect explicitly. Negative impressions from female students concentrated more on the organization of the units or on the level of complexity, and less on the app itself. On the other hand, male students mentioned app related aspects, like programming and Create@School's simplicity, as positive impressions significantly much more often than the units. Their dislikes significantly concentrated on the app, e.g., that it was buggy. This reflects the complaints of only six girls. Surprisingly, answers from girls also included that they did not like coding at all, whereas this was not an answer given by boys. The low performance and high error rate of the app is still a serious issue, but seems to have been more problematic for the male students. Possible reasons are again that boys are more used to utility apps ("tools") and game engines than girls are (Krieger, Allen, and Rawn, 2015).

To conclude, there are statistically significant different aspects which are more important for girls than for boys. In the literature review, the authors already described that CS lessons are mostly constructed to suit the interests of males. However, this evaluation shows that the working process and the sequence of the units are particularly important for engaging female students. Thus, for them not only is the tool essential but the learning environment as a whole, as well as the ability to express their own interests, e.g., through designing and creative activities. Based on these statistically significant results and our analysis of the existing literature, and after the completion of the NOLB project, we started to design more suitable learning environments for girls, particularly focussing on aspects of gender sensibility and awareness (McLean and Harlow, 2017). We shortly describe them in the next section and will publish a more detailed description in a future paper.

## Outlook

One result of NOLB, the Create@School app, was a great opportunity to provide schools with a tailored package of tools and services to help them integrate coding in their classrooms and to apply the app for interdisciplinary project work. This is important for the future to support teachers in Austria with the upcoming challenge to integrate a basic set of digital literacy education[3] in secondary schools. After NOLB, the authors concentrated their research on how to apply the NOLB results to tailor the Catrobat services to female needs. Therefore, we have developed a new course model that includes extrinsic and intrinsic motivators as well as four key fields which have been considered as important for coding activities: playfulness, engagement, creativity, and coding. This model has already been tested with two classes of 23 students (one mixed gender class and one girls-only class). This sample clearly was very small, and thus we plan further larger studies in the same and in different contexts (outside school). As a first step, an intensive "Girls Coding Week" will be performed later in 2018 to test the model's efficiency.

---

[3] Mandatory exercise "Digital literacy" in secondary education 1, Content for piloting in the school year 2017/18, [online] https://tinyurl.com/y78wov7a, accessed: 8.3.2018.